# Early completion based on multiple dosages to accelerate maximum tolerated dose-finding

Masahiro Kojima[1,2]

[1]Biometrics Department, R&D Division, Kyowa Kirin Co., Ltd., Tokyo, Japan.

[2]Department of Statistical Science, School of Multidisciplinary Sciences, The Graduate University for Advanced Studies, Tokyo, Japan.

**Running title**: Early Completion for Model-Assisted Designs

**Corresponding author**

Name: Mr. Masahiro Kojima, MSs

Address: Biometrics Department, R&D Division, Kyowa Kirin Co., Ltd.

Otemachi Financial City Grand Cube, 1-9-2 Otemachi, Chiyoda-ku, Tokyo, 100-004, Japan.

Tel: +81-3-5205-7200

FAX: +81- 3-5205-7182

Email: masahiro.kojima.tk@kyowakirin.com

**A conflict of interest disclosure statement**: None

**Word count**: 3229

**The number of figures**: 3

**The number of tables**: 1

## Abstract

**Background:** Phase I trials desire to identify the maximum tolerated dose (MTD) early and proceed quickly to an expansion cohort or phase II trial for efficacy. We propose an early completion method based on multiple dosages to accelerate the identification of MTD for model-assisted designs.

**Methods**: The early completion performs based on each dose-assignment probability for multiple dosages. The formulas of probabilities are simple to calculate. We evaluated the early completion for an actual trial. In addition, a simulation study performed the percentage of correct MTD selection and early completion.

**Results**: In the actual trial, the early completion could perform, and the MTD estimation phase could be shortened. In the simulation study, the change in the percentage of correct MTD selection from non-early completion version was 2.2% at the maximum, indicating that there was almost no decrease. In addition, depending on the toxicity rate, the percentage of correct MTD selection improved compared to the non-early completion version. The early completion performed from 18.1% to 90.8%.

**Conclusion**: We propose the early completion method with not only little reduction in accuracy but also improving the percentage of correct MTD selection. We conclude the early completion can perform unproblematically for model-assisted design.

## Introduction

The objective of phase I dose-finding trials is to identify the maximum tolerated dose (MTD) based on the occurrence of dose-limiting toxicities (DLTs). After identifying the MTD, an expansion cohort and/or a phase II trial to confirm efficacy start (1, 2, 3, 4, 5, 6, 7, 8, 9).

As a design aimed at estimating MTDs, the algorithm-based design represented by 3+3 design and the model-based design represented by CRM have been used. However, while the algorithm-based design is easy to operate, a performance is often insufficient. Although the model-based design has superior performance, its operation is complicated. Hence, a model-assisted design is proposed as a design combining a superior performance design such as a model-based design with a simple dose adjustment rule such as an algorithm-based design (10).

The model-assisted design cannot complete until the number of patients treated reaches the sample size. Although it is possible to set the number of patients treated to terminate a trial, the number is large and arbitrary (details in Supplemental Description 1). Therefore, it is important to identify MTD early with a small number of patients treated with sufficient accuracy using trial data. In addition, early identification of MTD allows to proceed to the expansion cohort or phase II trial more quickly.

In this paper, we propose a novel early completion method based on multiple dosages to accelerate the identification of MTD for model-assisted designs. By using not only the DLT data of the current dose but also the DLT data of higher and lower doses, the MTD can be identified early with almost no reduction in the percentage of correct MTD identification. We demonstrate the early completion to an actual trial. Simulation studies confirm the percentage of correct MTD selection and the performance of the early completion with 12 scenarios.

# Methods

## Model-assisted designs

Typical model-assisted designs are the modified toxicity probability interval (mTPI) design (11), the keyboard design (12) which is proposed to prevents an overdose in the mTPI design, and the BOIN design (13) which has the simplest dose-assignment rule. Various extensions of model-assisted design have also been proposed (14, 15, 16, 17, 18, 19, 20, 21). The early completion of model-assisted design has been proposed by Kojima (22,23). However, percentages of correct MTD identification with early completion decrease depending on DLT rates does because the early completion performs only based on the DLT data of current dose.

## Novel Early Completion based on Multiple dosages

We propose a novel early completion method based on multiple dosages (ECMD) for the model-assisted designs. We assume a phase I dose-finding trial that a sample size is $N$, a dose level is $J$, the current dose is $j$, the total number of patients treated at the current dose $j$ is $n_j$, the total number of DLTs at the current dose $j$ is $m_j$, the number of remaining patients is $r$. If the dose is not specified, $j$ is omitted. The ECMD performs based on three probabilities: a probability of not dose-de-escalation at the current dose $j$, a dose-de-escalation probability at the one higher dose $j+1$, and dose-escalation probability at the one lower dose $j-1$. To calculate the three probabilities, we consider the boundaries of dose-assignment. The maximum number of DLTs at the dose-escalation decision for $n_j+r$ patients at the current dose is defined as $E_{n_j+r}$. The $n_j+r$ patients refer to the maximum number of DLTs for which the dose escalation is determined when the current number of patients is $n_j$ is added the remaining patients $r$ that would be treated. The minimum number of DLTs at the dose de-escalation decision for $n_j+r$ patients at the current dose is defined as $D_{n_j+r}$. For example, on the BOIN design with a sample size of $N=21$, and 12 patients were treated. $n_j$ =6 patients received the current dose $j$, $m_j$ =2 patients presented with DLT, and $r=9$ patients remained. Therefore, $E_{n_j+r}=E_{6+9}=E_{15}=3$ and $D_{n_j+r}=D_{6+9}=D_{15}=6$ given by the dose-assignment table (Table 1).

The three probabilities to determine the ECMD are as follows,

**1.** dose-escalation probability at $j-1$: $BB\left(E_{n_{j-1}+r}-m_{j-1}; r, m_{j-1}, n_{j-1}\right)$

**2.** probability of not dose-de-escalation at $j$: $BB\left(D_{n+r}-1-m_j; r, m_j, n_j\right)$

**3.** dose-de-escalation probability at $j+1$: $1-BB\left(D_{n+r}-1-m_{j+1}; r, m_{j+1}, n_{j+1}\right)$

$BB(a; b, \alpha, \beta)$ is the cumulative beta-binomial distribution function that includes the number of successes $a$, the number of trials $b$, and the beta shape parameters $\alpha$ and $\beta$. The equation **1.** is the dose-escalation probability using the trial data $(m_{j-1}, n_{j-1})$ at the lower dose $j-1$ as the beta prior in the remaining $r$ patients, which refers to the probability that dose escalation is determined based on DLT in the $n_{j-1}+r$ patients treated. If the probability of equation **1.** is high, a probability is high that the MTD is within dosages higher than dose $j-1$. The equation **2.** is the probability of not dose-de-escalation using the trial data $(m_j, n_j)$ at the current dose $j$ as the beta prior in the remaining $r$ patients, which refers to the probability that dose-retainment or dose-escalation are determined based on DLT in the $n_j+r$ patients treated. Why the equation **2.** needs to include not only the dose-retainment probability but also the dose-escalation probability, the reason is to select the current dose as MTD even if a DLT rate is not high when the higher dose have a high DLT rate such that dose-de-escalation is assigned. The equation **3.** is the dose-de-escalation probability using the trial data $(m_{j+1}, n_{j+1})$ at the current dose $j+1$ as the beta prior in the remaining $r$ patients, which refers to the probability that dose-retainment or dose-escalation are determined based on DLT in the $n_{j+1}+r$ patients treated. If the probability is high, a probability is high that the MTD is within dosages lower than dose $j+1$.

The ECMD performs when all three probabilities exceed the threshold. The threshold value of 0.8 is recommended to ensure sufficient accuracy. Performances of the ECMD based on other thresholds are shown in the supplemental table 5. By using the three probabilities, when the higher dose shows a trend toward dose de-escalation and the lower dose shows a trend toward dose escalation, the ECMD is expected to be as accurate as non-early completion version. If the higher dose $j + 1$ has not been administered, no early completion decision is made. If the current dose is the highest dose, early completion is judged by excluding the dose-de-escalation probability of equation 3. If the current dose is the lowest dose, early completion is judged by excluding the dose-escalation probability of equation 1. If there is no DLT at the equations 1, 2, and 3, 0.5 is added to the conditional terms $n_{DLT}$ and $n$ from the cumulative beta-binomial distribution function. The rationale for adding 0.5 to $n_{DLT}$ and $n$ is described in Supplemental Description 2.

**Example trial**

The model-assisted designs were recently developed. Hence, there are no results from an active clinical trial. Using data from the completed clinical trials that used the continual reassessment method (CRM) similar to the model-assisted design, we demonstrate the EMCD for mTPI, Keyboard, and BOIN designs.

**PF-05212384 (PKI-587) trial (24) as an illustrative example**: The PKI-587 trial was the open label phase I trial using the modified CRM design for nonselected solid tumors. The primary objective was to evaluate safety. The part 1 of the trial was the MTD estimation phase. The planned eight dosages of veliparib were 10 mg, 21 mg, 43 mg, 89 mg, 154 mg, 222 mg, 266 mg, 319 mg, which were administered once weekly. The maximum sample size was 50. The planned cohort size was 2 to 4 patients. The target DLT level was 25%. The number of patients treated ($n$) and DLTs ($n_{DLT}$) at each dose were 10 mg ($n = 4,\ n_{DLT} = 0$), 21 mg ($n = 4,\ n_{DLT} = 0$), 43 mg ($n = 4,\ n_{DLT} = 0$), 89 mg ($n = 4,\ n_{DLT} = 0$), 154 mg ($n = 12,\ n_{DLT} = 0\ to\ 2$), 222 mg ($n = 7,\ n_{DLT} = 5$), 266 mg ($n = 8,\ n_{DLT} = 3$), and 319 mg ($n = 4, n_{DLT} = 2$). For the 154 mg, we were able to confirm that two DLTs occurred in a total of 42 patients, including 30 patients in the MTD confirmation phase, but we were unable to confirm how many MTDs occurred among the 12 patients evaluated in the MTD estimation phase. The ECMD will be evaluated with DLTs of 0, 1, and 2, respectively. Since DLTs occurred frequently at 222 mg, 154 mg as the current dose was used to perform the ECMD. We consider a situation where the dose was increased after the initial 154 mg dose, and the dose was reduced to 154 mg again due to a high DLT rate at the higher dose. Hence, in the situation after 8 patients have received 154 mg, we try the early completion. There are seven remaining patients who have not received any dose. The threshold for the ECMD is

0.8.

[*Case of* ***0*** *DLT out of 8 patients at 154 mg*]

For the mTPI design, the dose escalation probability at 89 mg is 0.82, the probability of not dose-de-escalation at 154 mg 1.00, and the dose-de-escalation probability at 222 mg is 1.00. The probabilities of all three are above the 0.8 threshold. Hence, the trial halts, and the MTD is identified following the rule of the design. For the Keyboard and BOIN designs, the dose escalation probability at 89 mg is 0.92, the probability of not dose-de-escalation at 154 mg 1.00, and the dose-de-escalation probability at 222 mg is 1.00. The probabilities of all three are above the 0.8 threshold. Hence, the trial halts, and the MTD is identified following the rule of the design.

[*Case of* ***1*** *DLT out of 8 patients at 154 mg*]

Since the probabilities at 89 mg and 222 mg remain the same, only the probability at 154 mg is shown. For the mTPI design, the probability of not dose-de-escalation at 154 mg is 0.99. The probabilities of all three are above the 0.8 threshold. Hence, the trial halts, and the MTD is identified following the rule of the design. For the Keyboard and BOIN designs, the probability of not dose-de-escalation at 154 mg is 0.97. The probabilities of all three are above the 0.8 threshold. Hence, the trial halts, and the MTD is identified following the rule of the design.

[*Case of* ***2*** *DLT out of 8 patients at 154 mg*]

Since the probabilities at 89 mg and 222 mg remain the same, only the probability at 154 mg is shown. For the mTPI design, the probability of not dose-de-escalation at 154 mg is 0.92. The probabilities of all three are above the 0.8 threshold. Hence, the trial halts, and the MTD is identified following the rule of the design. For the Keyboard and BOIN designs, the probability of not dose-de-escalation of 154 mg is 0.81. The probabilities of all three are above the 0.8 threshold Hence, the trial halts, and the MTD is identified following the rule of the design.

Early completion could perform for all DLTs 0, 1, and 2. We could reduce the number of patients treated by seven relative to the planned sample size.

## Numerical study

**Simulation settings.** We demonstrate the performance of the ECMD for the mTPI, Keyboard, and BOIN designs via a computer simulation using Monte Carlo experiment. In order to compare the performance evaluation of the ECMD, the non-early completion version and the early completion (EC) version proposed by Kojima (23) are evaluated in the same way. We prepare eleven designs: 3+3, CRM, mTPI, mTPI using the EC (mTPI-EC), mTPI using the ECMD (mTPI-ECMD), Keyboard, Keyboard using the EC (Keyboard-EC), Keyboard using the ECMD (Keyboard-ECMD), BOIN, and BOIN using the EC (BOIN-EC), and BOIN using the ECMD (BOIN-ECMD). We assume that the planned maximum sample size is 36, the dose level is six, and the target DLT level is 30%. The number of simulations times is 10,000. The threshold of ECMD is 0.8. The DLT probability at each dose is based on ten fixed scenarios and two random scenarios. The random scenarios are randomly set the DLT probability at each dose and the correct MTD to eliminate arbitrariness. Detailed descriptions of the fixed scenarios and random scenarios are provided in Supplemental Tables 1 and 2. Fixed scenario 1 assumes a case where less DLT is expected, as is observed with target therapies. The objective of fixed scenario 1 is to determine whether the trial is completed early. In fixed scenarios 2-7, the correct MTDs are set in order from the maximum dose to the minimum dose. Fixed Scenarios 8, 9, and 10 are the case when the correct MTD is not in the neighborhood of the target DLT level. Detail descriptions of each design are provided in Supplemental Description 3. We evaluated each method using the following criteria.

**Evaluation criteria.**

1. The percentage of correct MTD selection (PCMS)
2. Percent change from non-early completion version in the PCMS
3. Early completion percentage

The definition of percent change from non-early completion version in the PCMS is (PCMS of early completion − PCMS of non-early completion)/PCMS of non-early completion×100.

## Results

**Performance for the selection of the correct MTD.** Figure 1 shows the percentage of the correct MTD selection (PCMS) for the ten fixed scenarios and two random scenarios. The percentages for MTD selection for each dose of all designs are summarized in Supplemental Table 3 and 4. The PCMSs of mTPI-ECMD, Keyboard-ECMD, and BOIN-ECMD are close to the PCMSs of non-early completion version. The PCMSs of mTPI-EC are approximately 5% to 10% lower than the non-early completion version for all scenarios except for fixed scenario 1. For Keyboard-EC and BOIN-EC, PCMSs are almost the same as the non-early completion version in fixed scenarios 1, 3, 7, 8, and random scenario 2. However, the PCMSs of Keyboard-EC and BOIN-EC drop by roughly 4% to 5.5% in fixed scenarios 2, 4, 5, 6, 9, 10, and random scenario 1. The PCMSs of CRM are higher than the PCMSs of model-assisted designs in fixed scenario 1 and fixed scenarios 2, 4, 5, 6, 8, and 9 with correct MTD in medium dosage. In addition, PCMSs of CRM are also higher than the PCMSs of the model-assisted designs in random scenarios 1 and 2. However, in fixed scenarios 3, 7, and 10, where a prior information in the DLT response model of CRM deviated from the correct DLT rate, the performance is lower than in the model-assisted designs. The PCMSs of the 3+3 design are lower than the PCMSs of the other designs except for fixed scenario 7. The 3+3 design tends to select a lower dose as the MTD than the correct MTD. Hence, the PCMS of the 3+3 design is highest in fixed scenario 7 where the correct MTD is the lowest dose.

**Percent change from non-early completion version in the correct MTD selection.** Figure 2 shows the percent change from non-early completion version to the EC version and ECMD version in the correct MTD selection. The detailed values of the percent change are shown in Supplemental tables 3 and 4. mTPI-ECMD shows little change from -1.8% to 2.3%. The largest change -1.8% is only 0.7% decrease from 38.3% of mTPI to 37.6% in fixed scenario 3. The percent changes in Keyboard-ECMD ranges from -4.4% to 2.0%, with the largest change of -4.4% being a decrease of 2.2% from 49.5% to 47.4% in fixed scenario 8. The percent changes in BOIN-ECMD range from -4.2% to 2.0%, with the largest change of -4.2% being a decrease of 2.1% from 49.5% to 47.4% in fixed scenario 8. mTPI-EC ranges from -25.8% to 2.4%, the largest change of -25.8% was in fixed scenario 3, from 38.3% to 28.4%, a decrease of 9.9%. The percentage change in Keyboard-EC ranged from -8.4% to 2.4%. The largest change of -8.4% was in fixed scenario 2, from 48.7% to 44.6%, a decrease of 4.1%; the change in BOIN-EC ranged from -10.6% to 2.4%. The largest change in BOIN-EC, -10.6%, was in fixed scenario 5, from 52.0% to 46.5%, a decrease of 5.5%.

**Performance for early completion.** Figure 3 illustrates the percentages of early completion. The percentages of early completion of mTPI-ECMD range from 18.1% to 90.8%. Keyboard-ECMD and BOIN-ECMD are almost identical, ranging from about 29% to 98%. mTPI-EC ranges from 92.5% to 99.9%. Keyboard-EC and BOIN-EC range from 75% to 99%.

## Discussion

We proposed a novel early completion method based on multiple dosages (ECMD) to accelerate the MTD-finding on model-assisted designs. We confirmed that the ECMD performed with little reduction in the percentage of the correct MTD selection (PCMS) compared to the non-ECMD version because the decision of early completion is based on the current dose and the one higher and one lower doses. In addition, depending on the DLT rates, we confirmed to increase the PCMS from non-early completion version to ECMD version.

In the actual clinical trial (PKI-587) (24), the early completion could perform. Hence, we could reduce the number of patients treated by seven relative to the planned sample size. The safety observation per patient in the trial was 28 days. The reduction of seven patients refers to shorten 8 months if each patient takes 1 month to enroll. The reduction of 8 months is important in drug development.

We conducted 10,000 simulations for each design of 12 scenarios and confirmed that the PCMSs of ECMD version were almost the same compared to the PCMSs of the non-early completion version. In addition, Supplemental Table 5 showed the results of different thresholds for ECMD. We compared between threshold 0.8 and 0.9. Supplemental Figure 1 illustrates the percent change of PCMS from non-early completion version adding the results of threshold 0.9. The PCMS using threshold 0.9 improves slightly for almost scenarios. However, the PCMSs of 0.9 worsened in scenario 8 from not only ECMD version of threshold 0.8 but also EC version. In addition, the percentages of early completion for threshold 0.9 ranged about 20% to 90% lower for all scenarios compared to threshold 0.8. We recommend the threshold of 0.8 because the PCMSs are stable and early completion is conducted well. For different thresholds of EC version, Kojima (23) confirmed the PCMSs and the early completion rates, PCMSs did not significantly improve and early completion decreased at higher threshold. The maximum percent change in PCSM from the non-early completion version to the ECMD version was about −4%. On the other hand, in EC version, the maximum decrease was 25.8%, with seven scenarios showing a decrease of more than 10% and ten scenarios showing a decrease of more than 5%. The percent change in PCMS of the ECMD version from the EC version showed a maximum improvement in PCMS of 24.5% in the mTPI design, 6.2% in the Keyboard design, and 7.0% in BOIN design shown in Supplemental Table 6 and Supplemental Figure 2. Although the percentages of early completion of ECMD version were inferior to that of the EC version, we confirmed that the ECMD was conducted well and the number of total patients treated is reduced. We may be interested in the efficacy data of patients treated in the MTD in the MTD estimation phase.

The number of patients treated in the MTD of ECMD version was higher than in the EC version. Hence, we can also confirm the efficacy data. Detailed results are in Supplemental Tables 3 and 4.

We recommend applying the ECMD to all mTPI, Keyboard, and BOIN designs based on the accuracy. However, since the mTPI design is known to have a high risk of overdose (12,25), the Keyboard or BOIN designs are better. In particular, we recommend the BOIN-EMCD design having the simplest dosing assignment rules.

**Acknowledgments**: The author thanks Associate Professor Hisashi Noma for his encouragement and helpful suggestions.

**Author's Contributions**

M. Kojima: Conception and design; development of methodology; acquisition of data (provided animals, acquired and managed patients, provided facilities, etc.): analysis and interpretation of data (e.g., statistical analysis, biostatistics, computational analysis); writing, review, and revision of the manuscript; administrative, technical, and material support (i.e., reporting and organizing data, constructing databases); and study supervision.

**Table 1.** Dose escalation and de-escalation boundaries (TTL=0.3)

| Design | Action | Num of patients treated at current dose | | | | | |
|---|---|---|---|---|---|---|---|
| | | 3 | 6 | 9 | 12 | 15 | 18 |
| mTPI | Escalate if Num of DLTs $\leq$ | 0 | 1 | 1 | 2 | 2 | 3 |
| | De-escalate if Num of DLTs $\geq$ | 2 | 3 | 4 | 5 | 7 | 8 |
| Keyboard | Escalate if Num of DLTs $\leq$ | 0 | 1 | 2 | 2 | 3 | 4 |
| | De-escalate if Num of DLTs $\geq$ | 2 | 3 | 4 | 5 | 6 | 7 |
| BOIN | Escalate if Num of DLTs $\leq$ | 0 | 1 | 2 | 2 | 3 | 4 |
| | De-escalate if Num of DLTs $\geq$ | 2 | 3 | 4 | 5 | 6 | 7 |

**Figure 1.** Percentage of correct MTD selection for the all scenarios. EC: Design using early completion proposed by Kojima (23). ECMD: Design using novel early completion based on multiple dosages proposed in this paper.

[Fixed Scenario 1]

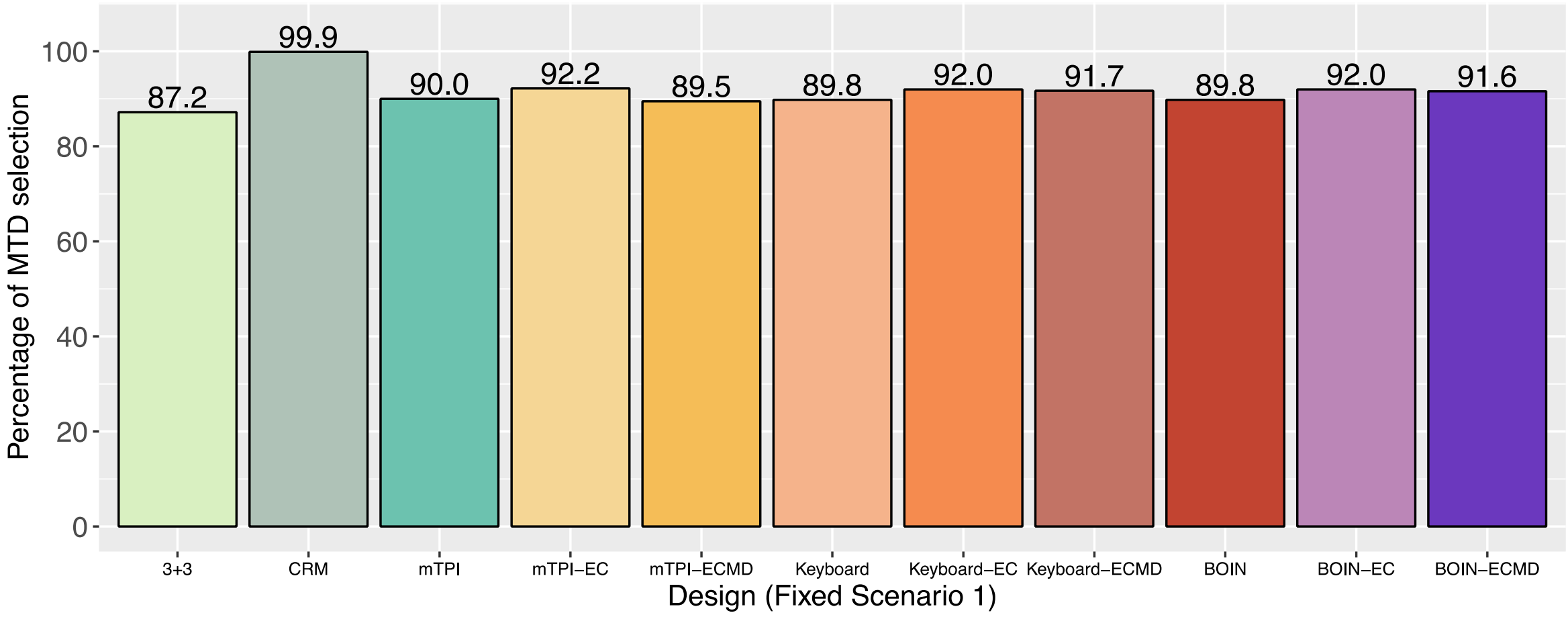

[Fixed Scenario 2]

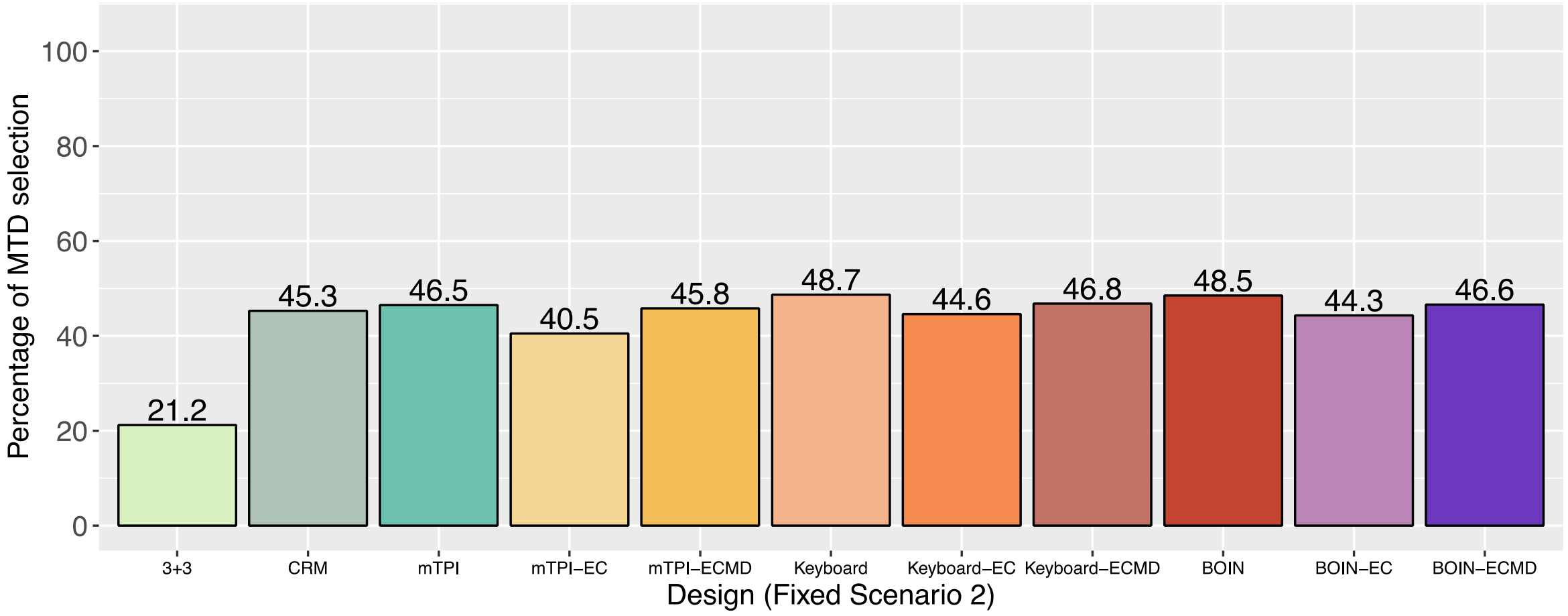

[Fixed Scenario 3]

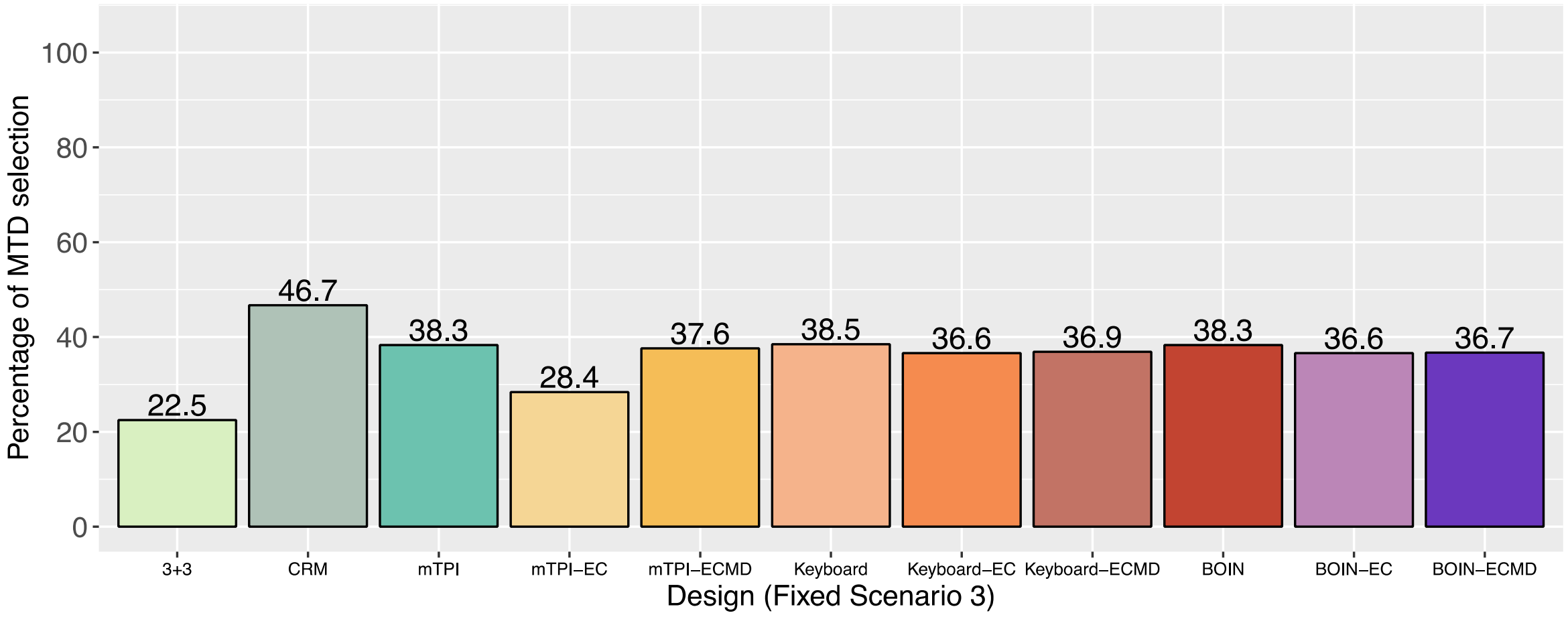

[Fixed Scenario 4]

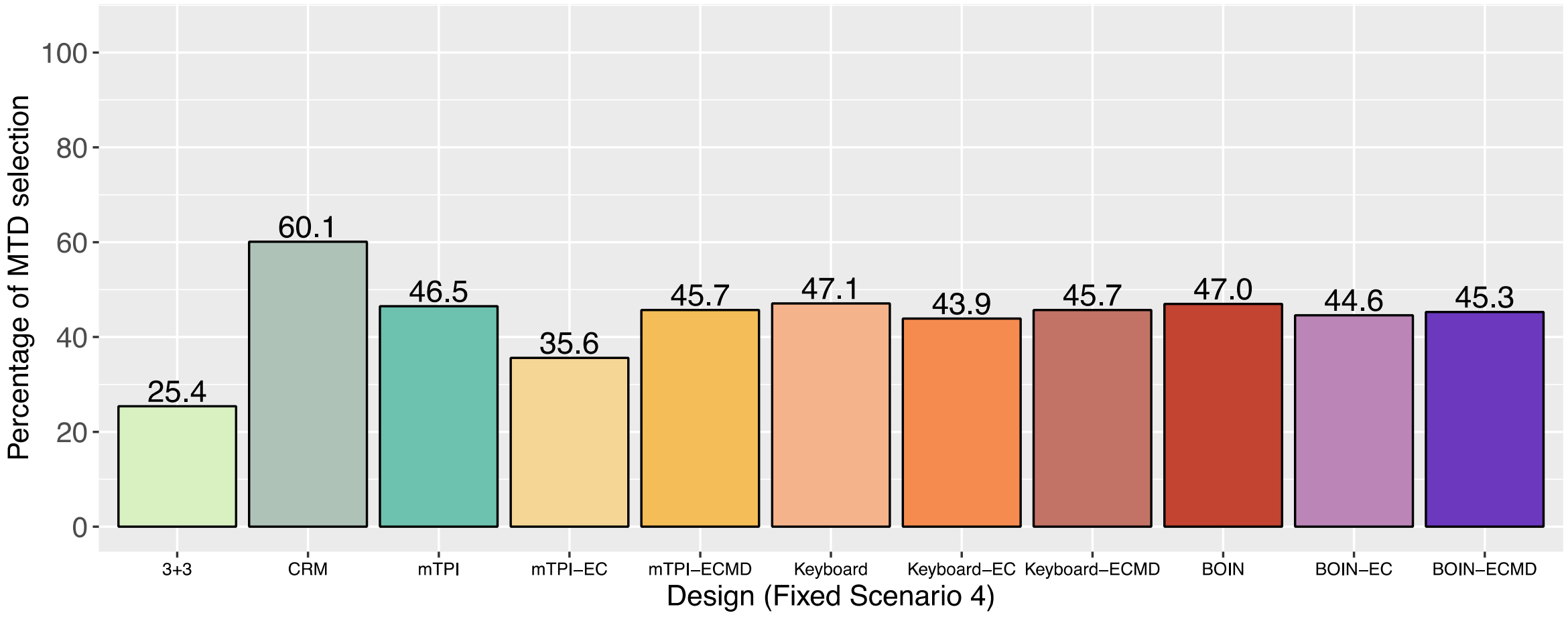

[Fixed Scenario 5]

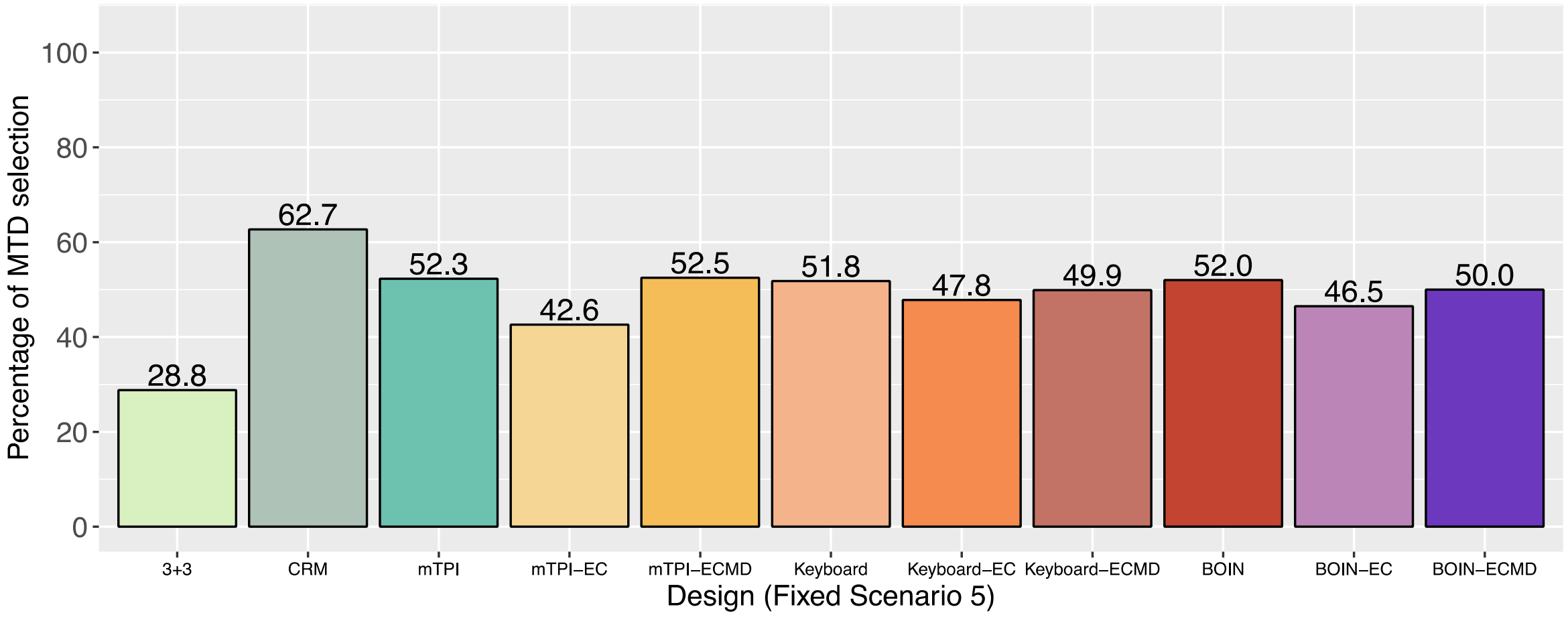

[Fixed Scenario 6]

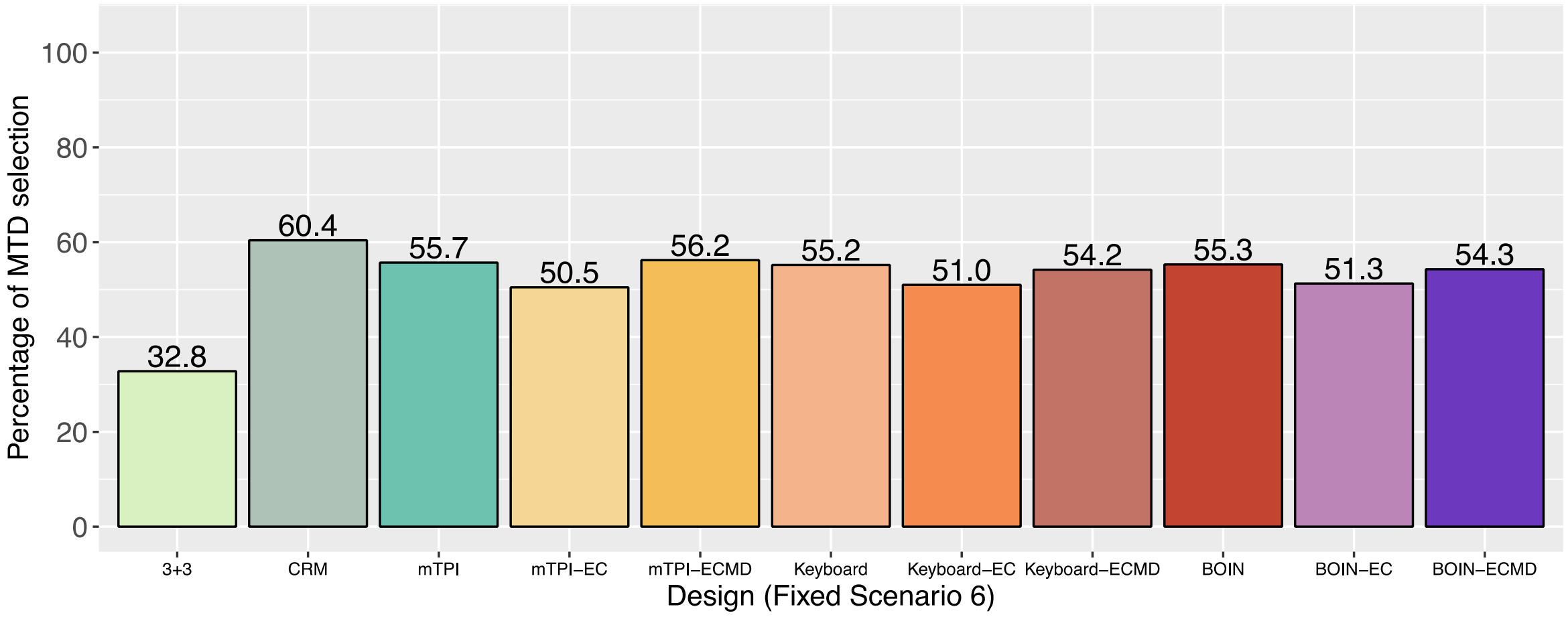

[Fixed Scenario 7]

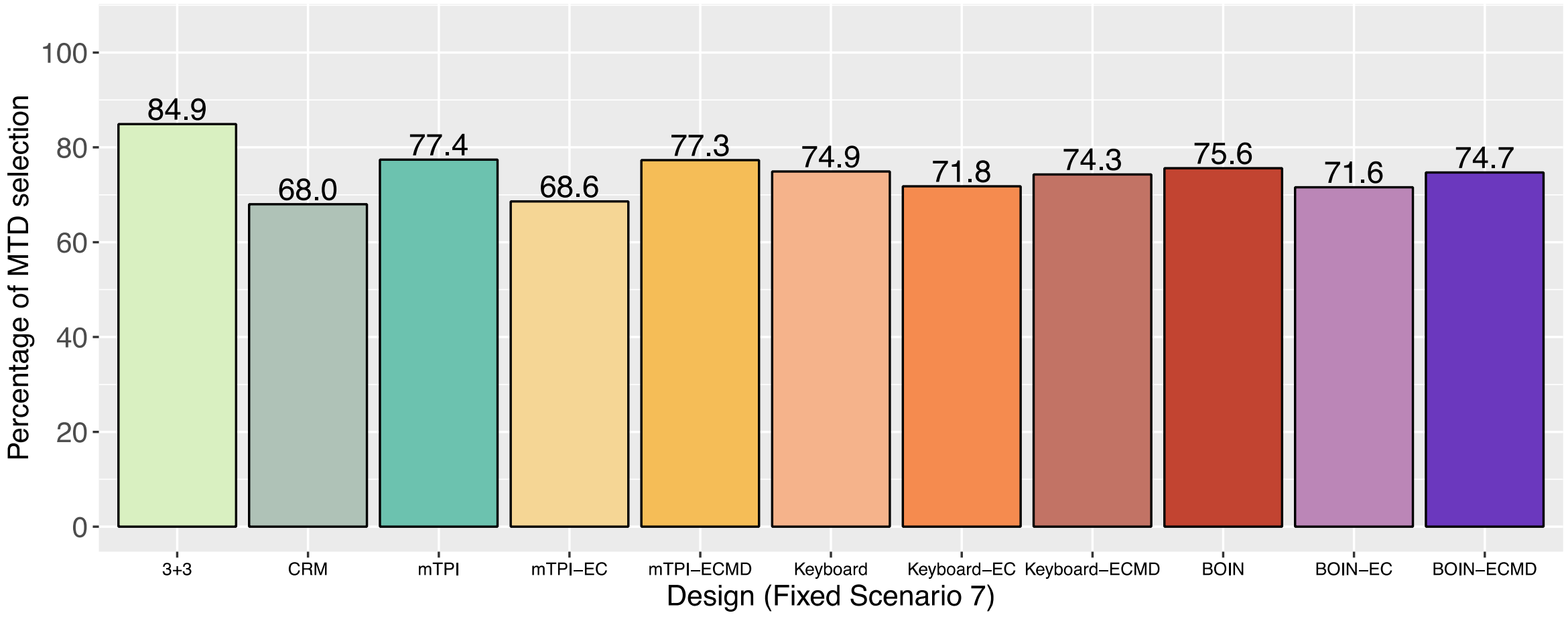

[Fixed Scenario 8]

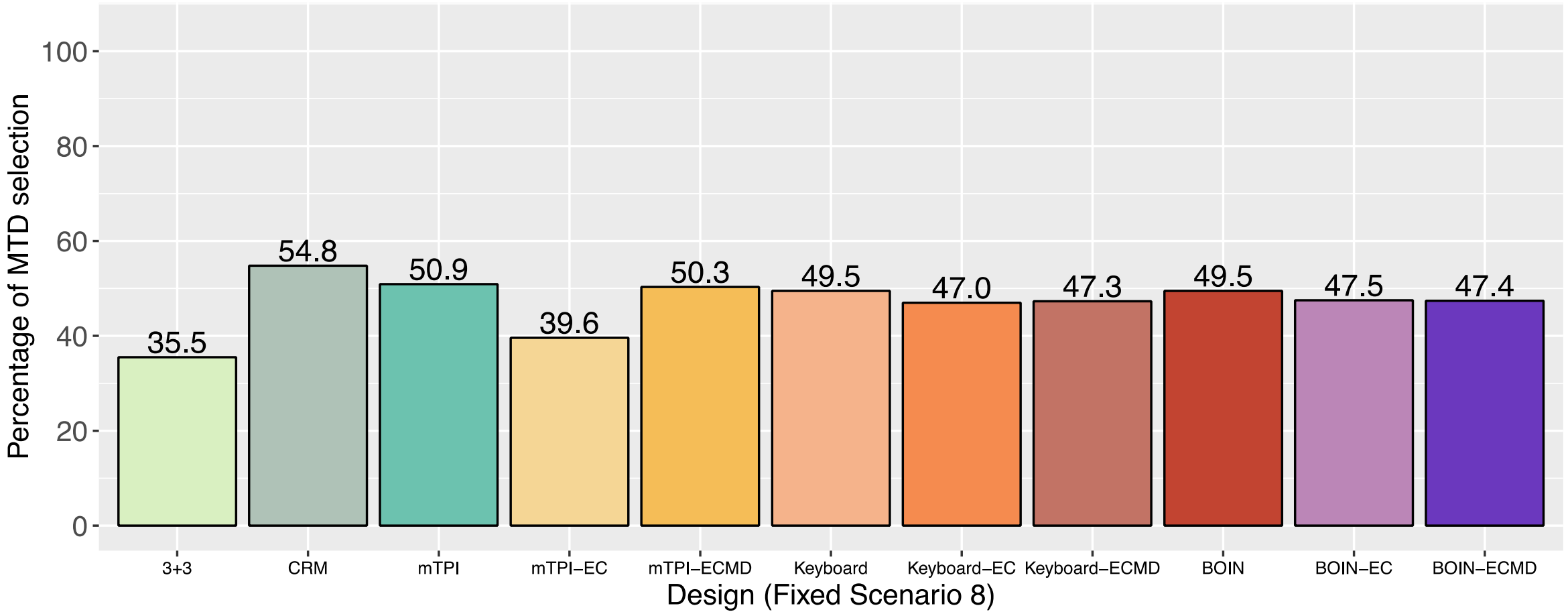

[Fixed Scenario 9]

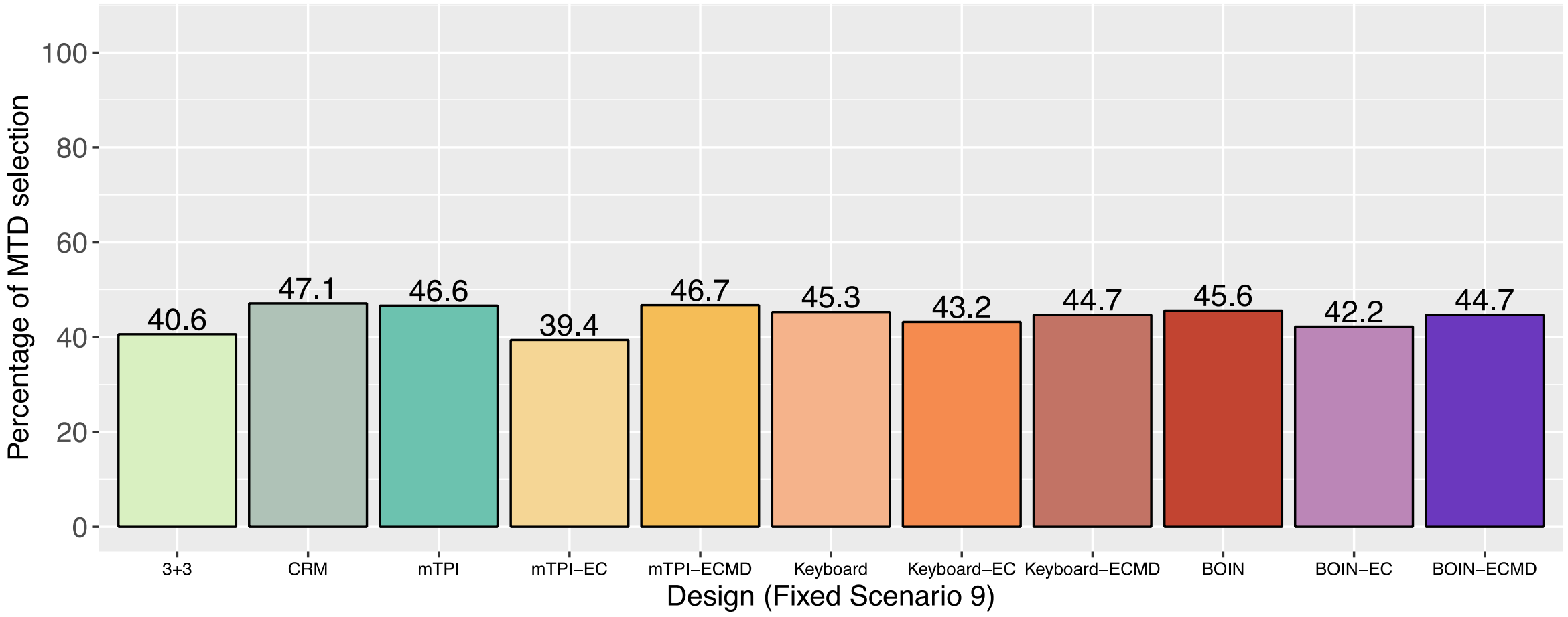

[Fixed Scenario 10]

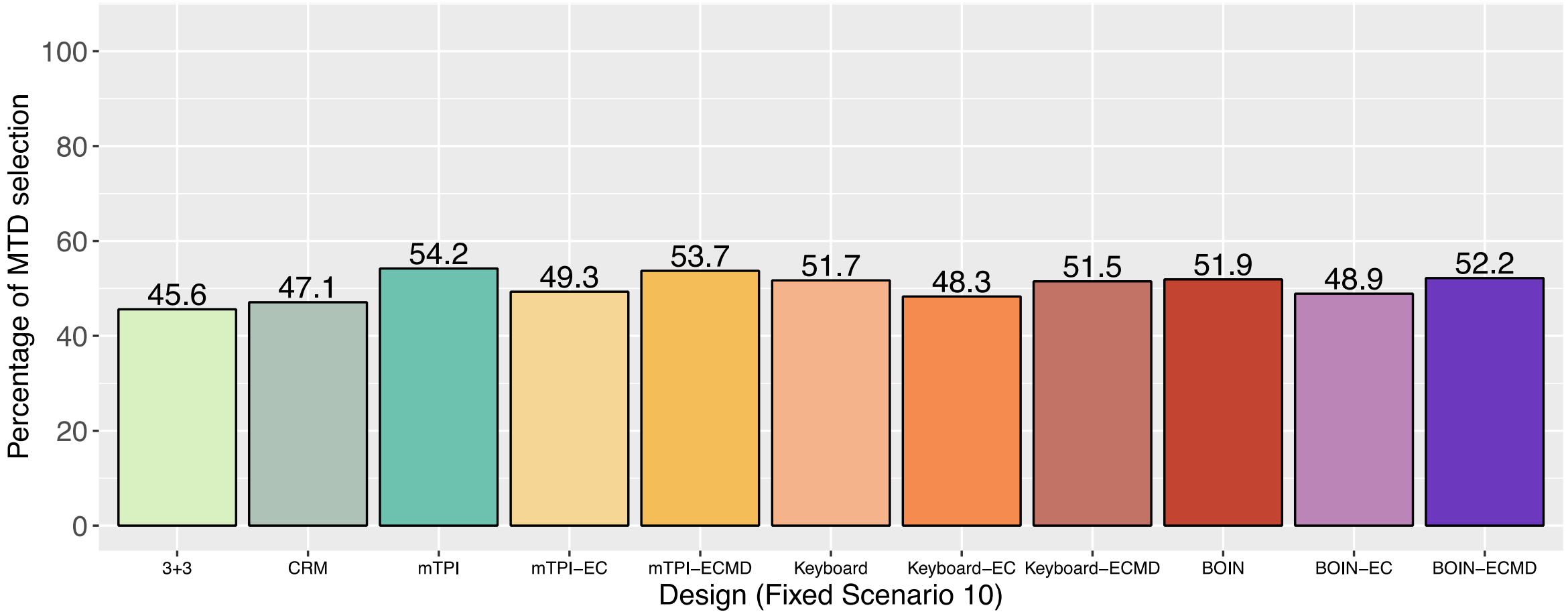

[Random Scenario 1 ($\mu = 0.5$)]

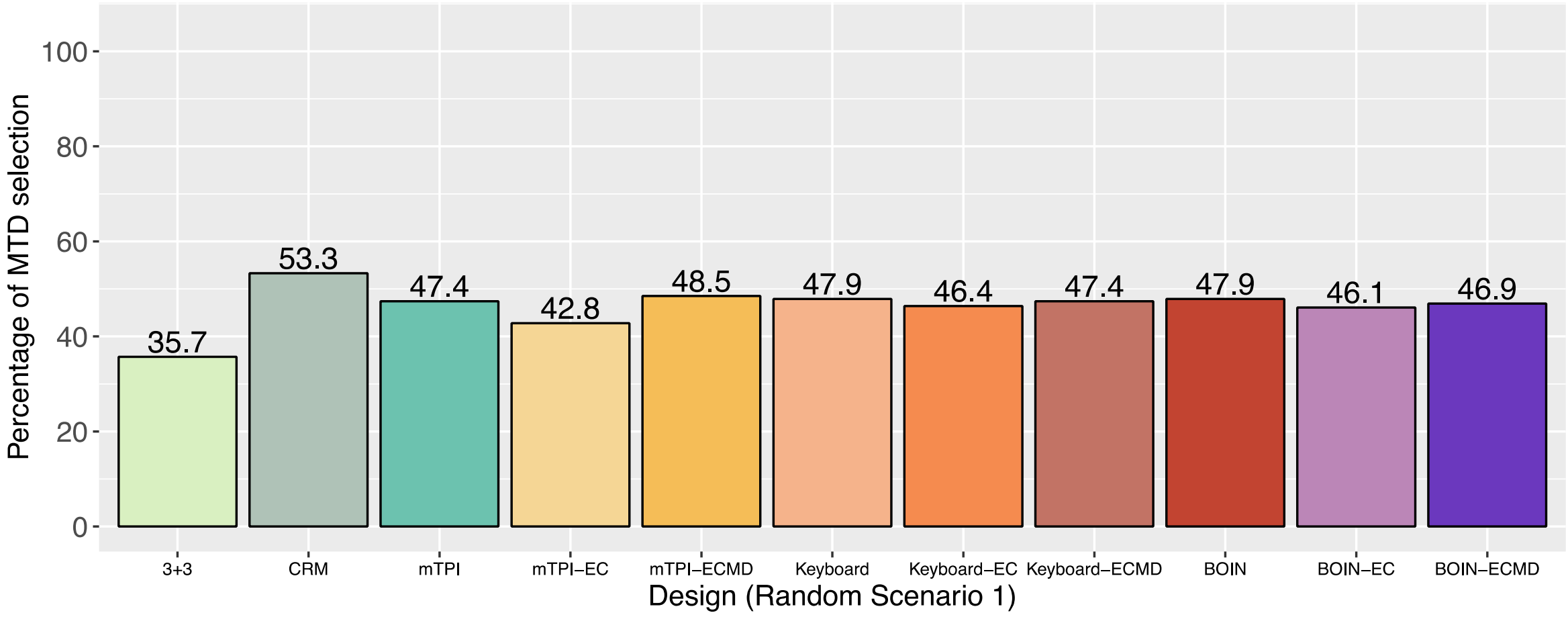

[Random Scenario 2 ($\mu = 1.0$)]

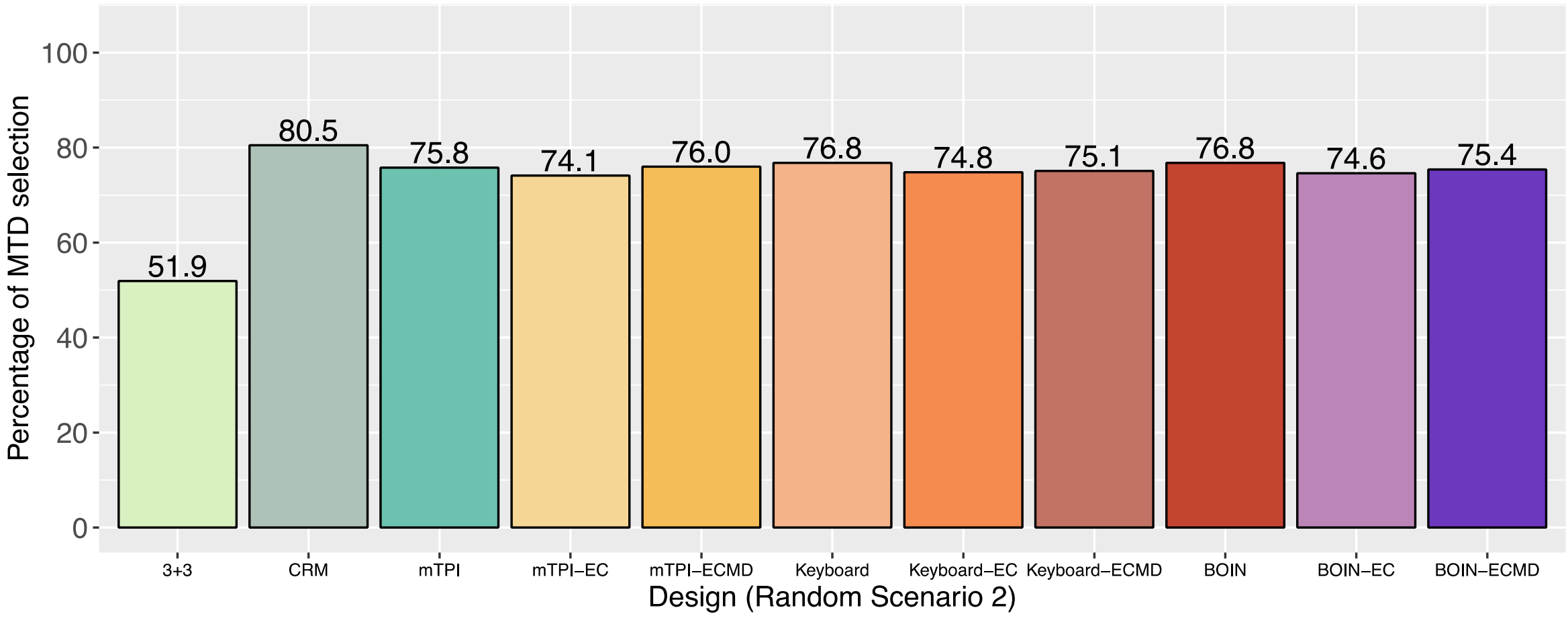

**Figure 2.** Percent change from non-early completion version in correct MTD selection

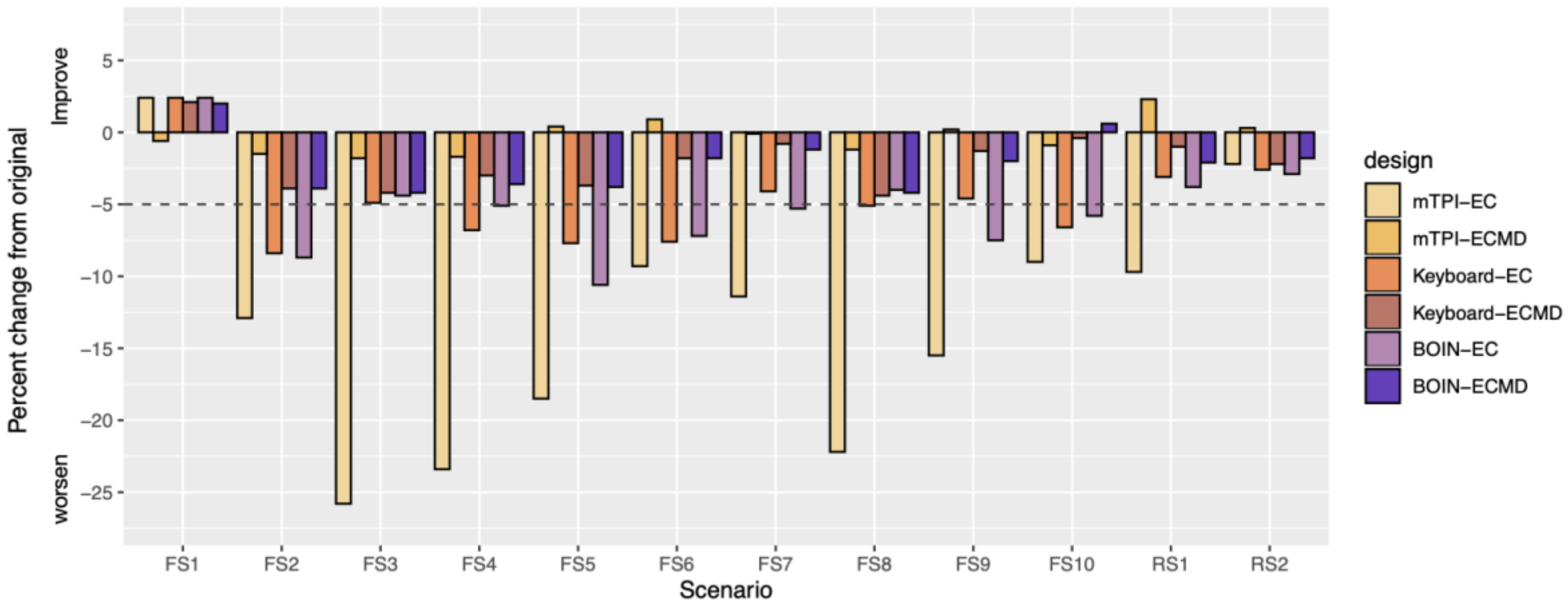

FS1: Fixed Scenario 1; FS2: Fixed Scenario 2; FS3: Fixed Scenario 3; FS4: Fixed Scenario 4; FS5: Fixed Scenario 5; FS6: Fixed Scenario 6; FS7: Fixed Scenario 7; FS8: Fixed Scenario 8; FS9: Fixed Scenario 9; FS10: Fixed Scenario 10; RS1: Random Scenario 1; and RS2: Random Scenario 2. EC: Design using early completion proposed by Kojima (23). ECMD: Design using novel early completion based on multiple dosages proposed in this paper.

**Figure 3.** Percentage of early completion

[Percentage of early completion for the all scenarios]

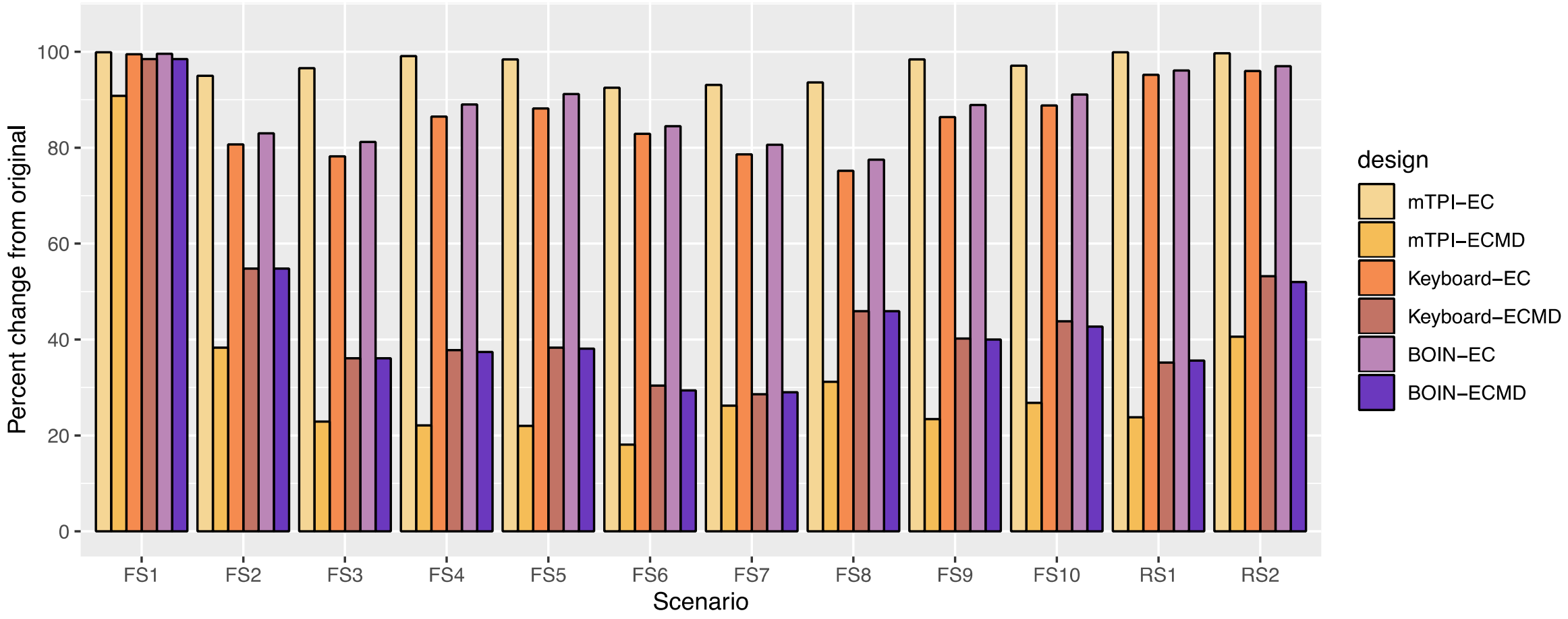

FS1: Fixed Scenario 1; FS2: Fixed Scenario 2; FS3: Fixed Scenario 3; FS4: Fixed Scenario 4; FS5: Fixed Scenario 5; FS6: Fixed Scenario 6; FS7: Fixed Scenario 7; FS8: Fixed Scenario 8; FS9: Fixed Scenario 9; FS10: Fixed Scenario 10; RS1: Random Scenario 1; and RS2: Random Scenario 2. EC: Design using early completion proposed by Kojima (23). ECMD: Design using novel early completion based on multiple dosages proposed in this paper.